\documentclass[aps,prb,twocolumn,showpacs,amsmath,amssymb]{revtex4-2}

\usepackage{graphicx}
\usepackage{xcolor}
\usepackage{bm,upgreek}
\usepackage{amsmath}
\usepackage[normalem]{ulem} 

\begin{document}
\title{Deeper insight into the terahertz response of conventional superconductors under magnetic field}

  \author{M. \v{S}indler$^1$}
\email{sindler@fzu.cz}
 \author{F.~Herman$^2$}
  \author{F.~Kadlec$^1$}
  \author{C.~Kadlec$^1$}
\email{kadlecch@fzu.cz}

\affiliation{$^1$Institute of Physics, Czech Academy of Sciences, Na Slovance 2, 182 21 Prague 8, Czech Republic}
\affiliation{$^2$Department of Experimental Physics, Comenius University, Mlynská Dolina~F2, 842 48 Bratislava, Slovakia}
  \date{\today}




  \begin{abstract}
 We investigate the terahertz conductivity of conventional superconductors in Voigt and Faraday magneto-optical configurations.
  First, we review theoretical approaches describing the fundamental processes of suppression of superconductivity in magnetic field and how the in-gap states are filled.
 In the Voigt geometry, thin superconducting films are fully penetrated by the magnetic field which interacts with the spin, thus modifying the magnitudes of the optical gap and of the density of the condensate. In this configuration, we provide an alternative description of the recent experiments showing the gapless conductivity of a Nb film measured by Lee {\it et al.} [Nat. Commun. \textbf{14}, 2737 (2023)], which better fits their data for magnetic fields above 1 T. In the Faraday geometry, we measured and analyzed the terahertz conductivity of three NbN films with varying thicknesses using the Maxwell-Garnett model, treating vortices as normal-state inclusions within a superconducting matrix. 
 In both geometries, the optical conductivity can be comprehensively described by the model of Herman and Hlubina [Phys. Rev. B \textbf{96}, 014509 (2017)] involving pair-conserving, and magnetic-field-dependent pair-breaking disorder scattering processes.
 
  \end{abstract}



  \maketitle

\section{Introduction}
\label{intro}

Magnetic field has dramatic effects on super\-con\-duc\-ti\-vity---it induces screening currents and the appearance
of a vortex lattice (orbital effect) and it interacts with
the electron spin (Zeeman effect). The contributions of
these two effects to the terahertz (THz) conductivity depend on
the orientation of the applied magnetic field with respect to the sample (see Fig. 1).
In the Faraday geometry, the magnetic field is applied perpendicular to the surface, which results in screening currents suppressing the field inside the superconductors, thus the orbital effect usually dominates. In the Voigt geometry, the magnetic field is directed along the surface. In films with a thickness smaller than the penetration depth, the orbital effect becomes weaker with decreasing film thickness. The Zeeman effect thus dominates in ultrathin samples~\cite{Meservey1970}.  

\begin{figure}
\centering
  \includegraphics[width=0.45\textwidth]{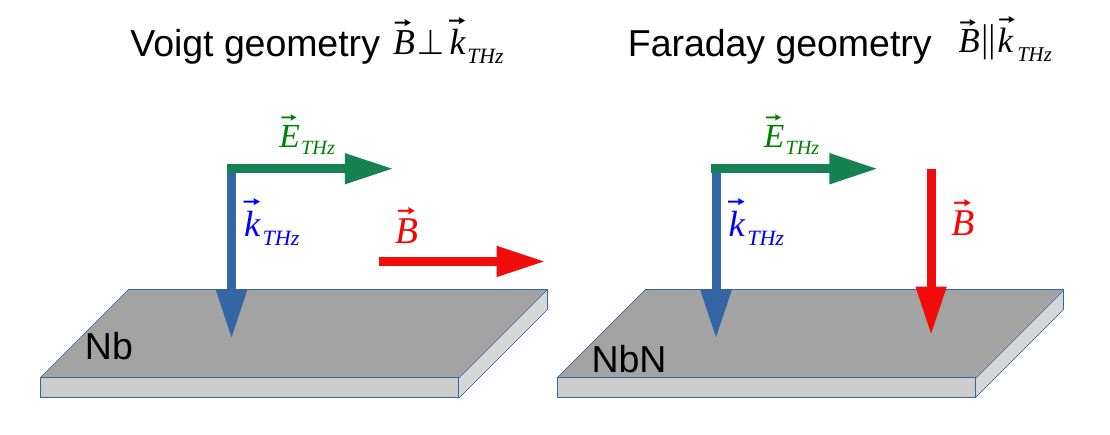}
\caption{Magneto-optical geometries. $\vec{B}$: external magnetic field. The THz beam propagates perpendicular to the surface of the sample, with a wave vector $\vec{k}_{\mathrm{THz}}$ and an electric field $\vec{E}_{\mathrm{THz}}$.}
\label{scheme}       
\end{figure}

Abrikosov developed a theoretical approach able to account for the pair-breaking effect of paramagnetic impurities~\cite{Abrikosov1961}. 
 Later, Maki~\cite{Maki1963} showed that the Abrikosov approach
could be used for other pair-breaking mechanisms including external
magnetic field. Finally, Skalski, Bet\-be\-der-Ma\-ti\-bet and
Weiss (SBW) worked out an integral expression
for optical conductivity that accounts for pair-breaking effects~\cite{Skalski1964}. The SBW model was recently used to explain the THz spectra
of niobium, NbN and  Nb$_{0.5}$Ti$_{0.5}$N films in the Voigt geometry \cite{Lee2023,Xi2010,Xithesis}. Xi \textit{et al.}~\cite{Xi2010,Xithesis} observed the suppression of
the pair-cor\-re\-la\-tion gap $\Delta$ and of the spectroscopic gap $\Omega_G$  in
ultrathin films of NbN and Nb$_{0.5}$Ti$_{0.5}$N in agreement
with the SBW model~\cite{Skalski1964}, using the pair-breaking parameter $\alpha$
to account for the effects of the magnetic field. While for NbN, 
$\alpha $ is linearly proportional to $B$ ($\alpha \sim  B$),
for Nb$_{0.5}$Ti$_{0.5}$N, a square law has been observed ($\alpha \sim B^2$).

In similar measurements on a thin niobium film~\cite{Lee2023}, for a temperature of 1.5~K, Lee \textit{et al.} observed a complete suppression of the spectroscopic gap at 2.4~T. Between 2.4 and 3.5~T, $\Omega_G=0$ while $\Delta$ was still non zero. In magnetic fields above 3.5~T, the superconductivity was lost. These THz spectra were well described by the SBW model even in the gapless regime. However, in contrast to the observed data, the SBW model predicts a steep fall in the real part of conductivity near low frequencies for magnetic fields of 1--2 T. 

In this paper, we propose a way of
interpreting the THz spectra in magnetic field by effectively replacing
the SBW model with the one developed by Herman and
Hlubina (HH model)~\cite{Herman2017} which includes a heuristic dependence of the pair-breaking parameter on the external magnetic field, in a similar fashion as in Ref.~\cite{Sindler2022}. Notice that the HH model has basically the same number of parameters as the SBW model. This makes them equally computationally challenging and allows for a simple comparison of their results. Moreover, the resulting analytical formulation of the HH model using a Green's function allows for a more straightforward insight into several physical quantities, including the density of states (DOS).

In the Faraday geometry, there are two approaches to describing the THz conductivity. In the first one, vortex cores are treated as normal-state inclusions within a superconducting environment and this inhomogeneous system can be described as a homogeneous one with an effective complex conductivity. Maxwell-Garnett theory~\cite{Garnett1904} is the most commonly used~\cite{Xi2013,Sindler2014} as it correctly captures the topology of isolated vortex cores. In this approach, vortices are static and possible vortex motion and related consequences are thus neglected. In the second approach, the two-fluid model is utilized in various theoretical models~\cite{CCmodel1991,Brandt1991,Dulcic1993,Lipavsky2012} to describe the vortex dynamics as the dominant contribution to both the real and the imaginary parts of conductivity. The inherent weakness of the two-fluid model is that it does not account for the presence of the superconducting gap in the spectra, which can be justified for frequencies well below the gap but fails in the vicinity of the gap and for higher frequencies.
In this article, we propose improvements of the Xi's approach~\cite{Xi2013}, 
where the complex conductivity of the superconducting matrix is not described by its zero-field value but the HH model is used to account for the presence of the pair-conserving and pair-breaking scattering processes, the latter gaining in strength in the presence of external magnetic field.
\section{Theory}
\label{sec:theor}
\subsection{Fundamental aspects of the SBW and the HH models}

First, the DOS needs to be examined. The BCS theory~\cite{BCS} postulates the oversimplified formula 
        \begin{equation}
        \mathrm{DOS}= \Re \,\left( \frac{u}{\sqrt{u^2-1}} \right),
                \label{SBW_DOS}
         \end{equation}
         where $ \Re$ stands for the real part of the complex value and the parameter $u=E/\Delta$ equals the energy normalized to the value of the gap where E=0 at Fermi level.  For energy within the interval $-\Delta < E < \Delta $ there are no states and the DOS diverges at $E=\pm \Delta$. The Zeeman effect leads to a shift in the DOS for up and down spins, thus their respective DOS are $2\mu_B B$ apart, where $B$ is the magnetic field and
           $\mu_B =  9.274 \times 10^{-24}$  J T$^{-1}$ denotes the Bohr magneton. However, spin-orbit scattering leads to mixing the states for spin up and down states. The spin-orbit scattering rate is proportional to the fourth power of the atomic number~\cite{Crow1974,Meservey1976}, thus the described shift can be clearly observed only in samples with atoms with a low atomic number such as Al films~\cite{Meservey1970}. Here, we study the response of Nb and NbN films in the magnetic field, thus the DOS for up and down spins are smeared into one broad peak. Further details can be found in the Fulde's review paper~\cite{Fulde1973}.

In the limit of strong spin-orbit scattering, the SBW model generalizes the DOS from the BCS theory, Eq.~(\ref{SBW_DOS}), and $u$ is evaluated from the following nonlinear equation:
\begin{equation}
u\Delta =E+ \alpha \frac{u}{\sqrt{u^2-1}}
\label{eq:uDelta}
\end{equation}
where $\alpha$ is the Maki's pair-breaking parameter. The pair-correlation gap $\Delta$ is both temperature and magnetic field-dependent. In zero magnetic field, its value corresponds to the gap value obtained from the optical measurements $\Omega_G$. However, with increasing magnetic field, or more generally with increasing $\alpha$, $\Omega_G(B)$ decreases faster than $\Delta(B)$. At a critical value of the magnetic field, the superconductor enters a so-called gapless state where $\Omega_G$ becomes zero whereas $\Delta>0$ thus allowing the persistence of the superconducting state. At even higher values of the magnetic field, $\Delta$ becomes zero and the superconductivity is finally fully suppressed.

In contrast, the Dynes formula~\cite{Dynes1978} accounts
for the DOS broadening by phe\-no\-me\-no\-logi\-cally introducing a broadening parameter $\Gamma$:
        \begin{equation}
        \mathrm{DOS}\equiv N(E)= \Re \, \left(\frac{E+i\Gamma}{\sqrt{(E+i \Gamma)^2 - \Delta^2}} \right).
        \label{Dynes_DOS}
         \end{equation}

 A finite value of $\Gamma$ introduces subgap states at the Fermi level due to pair-breaking effects. 
 The formula has been experimentally verified through tunneling spectroscopy measurements in multiple superconducting systems, notably NbN~\cite{Chockalingam2009,Sindler2014}, Al~\cite{Dynes1984}, and MoC~\cite{Szabo2016}. Furthermore, this approach has successfully described spectroscopic features observed in photoemission studies of YB$_6$~\cite{Souma2005}.

The microscopic origin of the Dynes formula was unclear for a long time, but in a recent paper~\cite{Herman2016}, it was shown that the formula given by Eq.~(\ref{Dynes_DOS}) is valid in systems with a pair-breaking classical disorder provided that the pair-breaking scattering probability distribution has a Lorentzian distribution with width $\Gamma$ within the Coherent Potential Approximation (CPA).

\begin{figure}[htpb]
    \includegraphics[width = 0.48\textwidth]{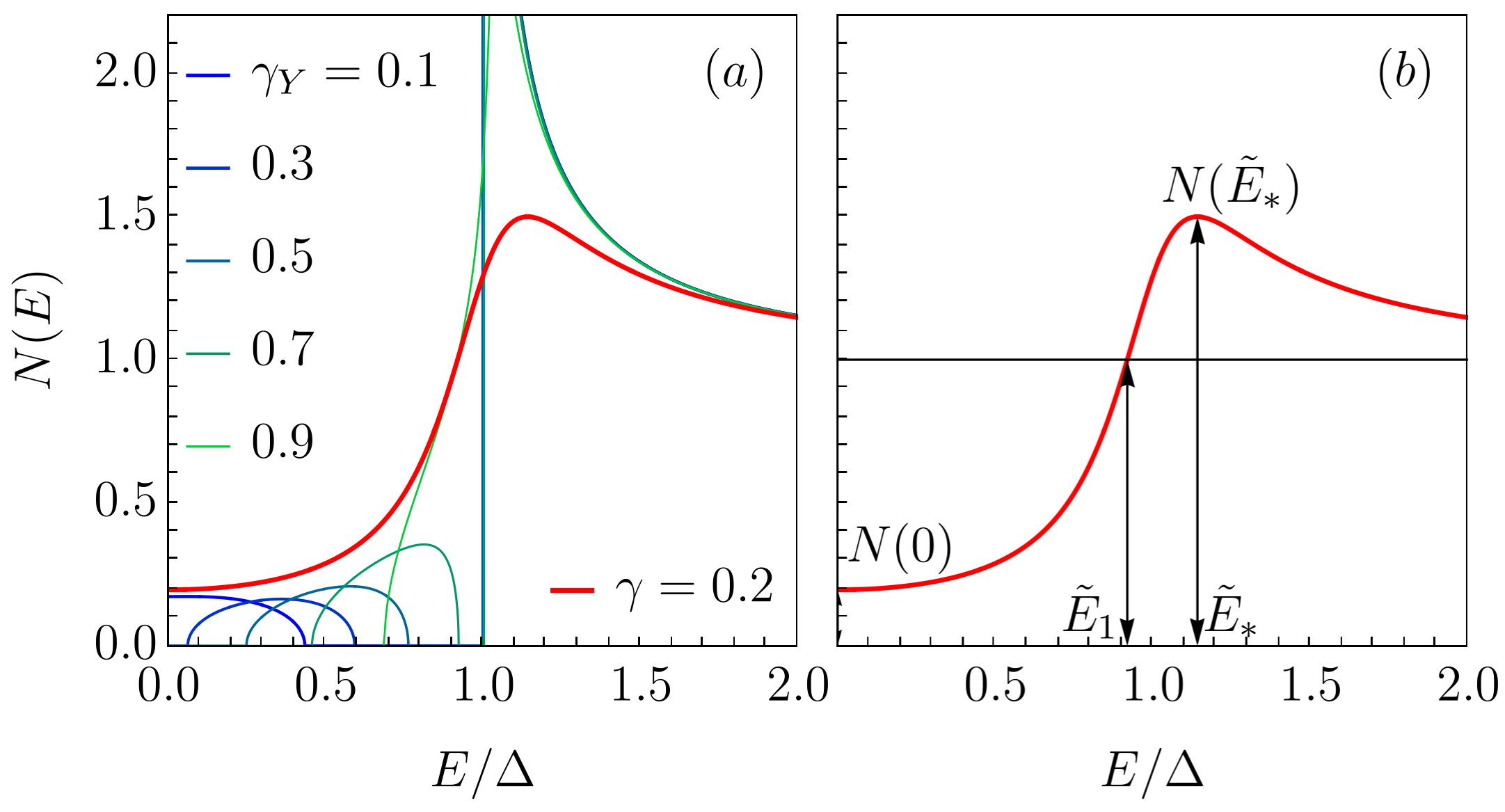}
    \caption{(a) $N(E)$ from the Dynes model (red curve) as an envelope function of different Yu-Shiba-Rusinov bound states ($\alpha = 0.04$). (b): The same Dynes $N(E)$ with important values marked.}
    \label{fig:dos_Dynes_Shiba}
\end{figure}

In Fig.~\ref{fig:dos_Dynes_Shiba}$(a)$ we plot the Dynes DOS together with curves corresponding to a DOS resulting from multiple scattering of the electron on a single magnetic impurity (assuming its low concentration), calculated within the T-matrix approximation \cite{Shiba68}. The described approach leads to the Yu-Shiba-Rusinov (YSR) sub-gap states in the superconducting gap of the resulting DOS. The center of the sub-gap states can be tuned by the YSR parameter $\gamma_Y$ \cite{Shiba68} depending on the coupling constant of the magnetic impurity. For simplicity, the value of the other YSR parameter $\alpha$ which expresses the amount of magnetic impurities, was taken as $\alpha=0.04$ in all cases. Notice that the Dynes DOS can be understood as an envelope function of the YSR impurity bands resulting from the multiple scatterings on magnetic impurities with different coupling strengths.

In Fig.~\ref{fig:dos_Dynes_Shiba}$(b)$ we indicate the important parameters of the Dynes DOS. The energy of the Dynes DOS maximum in the coherence peak amounts to
\begin{equation*}
    \tilde{E}_*^2 = 1 + 5\gamma^2/3+\sqrt{\left(1+5\gamma^2/3\right)^2-\left(1+\gamma^2\right)^2}
\end{equation*}
where $\tilde{E}_*=E_*/\Delta$ and $\gamma=\Gamma/\Delta = N(0)/\sqrt{1-N(0)^2}$. Note that if $N(0)$ is determined from tunneling measurements at very low temperatures, one can calculate $\gamma$ and $\tilde{E}_*$ which allows for a direct estimate of both the relevant gap $\Delta$ and the pair-breaking parameter $\Gamma$.

Also, another interesting energy scale $\tilde{E}_1$, the energy value when $N(\tilde{E}_1)=1$, can be also analytically expressed in the form
\begin{equation*}
    \tilde{E}_1^2 = 1-\frac{\gamma^2}{3}+\frac{u}{2^{1/3} 3}-\frac{2^{1/3}}{3}\frac{\gamma^2\left(15-4\gamma^2\right)}{u},
\end{equation*}
where
\begin{multline*}
u = \gamma^\frac{2}{3}\Big[16\gamma^4+153\gamma^2-27\\
+3\sqrt{3\left(27+194\gamma^2+435\gamma^4+288\gamma^6\right)}\Big]^{1/3}.
\end{multline*}

In the presented context, the parameter $\Gamma$ accounts for the influence of the magnetic (spin-flip)  scattering induced by the pair-breaking effect of the present magnetic field. In this way, we model the influence of Zeeman splitting, orbital effects, and spin-orbit interactions for the Voigt geometry and the presence of the screening currents in the Faraday geometry in a simple phenomenological approach. We also assume that the effective pair-breaking parameter $\Gamma(B)$ is an increasing function, thus playing a role similar to $\alpha$ in the SBW model. Moreover, the solution of the CPA equations \cite{Herman2016} results in convenient forms of the superconducting gap function $\Delta(E)$ and of the wavefunction renormalization $Z(E)$ \cite{Herman2017}:
\begin{eqnarray*}
    \Delta(E) &=& \overline{\Delta} \bigg/ \left(1 + \frac{i\Gamma}{E}\right),\\
    Z(E) &=& \left(1 + \frac{i\Gamma}{E}\right)\left(1 + \frac{i\Gamma_s}{\Omega(E)}\right),
\end{eqnarray*}
where \footnote{We choose the square root branch so that the signs of $\mathrm{Re}\{\Omega(E)\}$ and $E$ are the same.} $\Omega(E) = \sqrt{\left(E + i\Gamma\right)^2 - \overline{\Delta}^2}$ and $\overline{\Delta}=\overline{\Delta}(T)$ corresponds to the temperature-dependent gap \cite{Lebedeva2024a}. This formulation introduces the pair-breaking $\Gamma$ and the pair-conserving  $\Gamma_s$ scattering rates. It induces the Dynes tunnelling density of states and fulfills the Anderson theorem \cite{Anderson1959}. 
Although the main advantage of the HH model from our perspective is the clear identification of the pair-con\-ser\-ving and pair-breaking scatterings in the superconducting state, it also results in a natural normal-state limit $\Delta(E)=0$ and $Z_N(E) = (1+i\Gamma_N/E)$, where $\Gamma_N = \Gamma_s + \Gamma$ expresses the total effect of the disorder, as expected. In the normal state, the overall scattering rate is related to the scattering time $\tau$ in the Drude model for optical conductivity by $\Gamma_N=\hbar/ (2  \tau)$.

\subsection{Optical conductivity utilizing HH model}
The optical conductivity of conventional superconductors is calculated from the integral expression resulting from the Green function approach~\cite{Herman2017,Herman2016,Lebedeva2024b}.
Furthermore, its numerical evaluation is as costly as that which makes use of the generalized Mattis-Bardeen formula~\cite{Zimmermann1991} but, unlike the latter, it also allows for pair-breaking processes.

When discussing the optical response of conventional superconductors, it is customary to follow Nam \cite{Nam1967a} 
and introduce three complex functions: the density of states $n(\omega)$, the density of pairs $p(\omega)$ and a function $\epsilon(\omega)$ with a dimension of energy:
\begin{eqnarray}
n(\omega)&=&n_1+in_2
=\frac{\omega}{\sqrt{\omega^2-\Delta^2(\omega)}}=\frac{\omega+i\Gamma}{\Omega(\omega)},\label{eq:n(w)}\\
p(\omega)&=&p_1+ip_2
=\frac{\Delta(\omega)}{\sqrt{\omega^2-\Delta^2(\omega)}}=\frac{\overline{\Delta}}{\Omega(\omega)},\label{eq:p(w)}\\
\epsilon(\omega)&=&\epsilon_1+i\epsilon_2
=Z(\omega)\sqrt{\omega^2-\Delta^2(\omega)}=\Omega(\omega) + i \Gamma_s.\label{eq:e(w)}
\end{eqnarray}
The functions $n(\omega)$ and $p(\omega)$ are obviously
linked together, and they satisfy
$n^2(\omega) - p^2(\omega) = 1$. Note that the
functions $n_1(\omega)$, $p_2(\omega)$ and $\epsilon_2(\omega)$ are even, whereas $n_2(\omega)$, $p_1(\omega)$ and $\epsilon_1(\omega)$ are odd in $\omega$. 

Next, let us introduce auxiliary variables
\begin{eqnarray}
H_1 \big(x, y\big) &=&
\frac{1 + n(x)n^*(y) + p(x)p^*(y)}
{2\left[\epsilon^*(y) - \epsilon(x)\right]}, \nonumber\\
H_2 \big(x, y\big) &=& \frac{1 - n(x)n(y) - p(x)p(y)}
{2\left[\epsilon(y)+\epsilon(x)\right]}.
\label{eq:function_h}
\end{eqnarray}
Assuming the local limit of the electromagnetic field response, given the characteristic values of the considered mean free paths $\ell\sim 1\,\mathrm{nm}$, the resulting optical conductivity can be written as \cite{Herman2017,Lebedeva2024b}
\begin{eqnarray}
   \sigma_s(\omega)&=\frac{iD_0}{\omega}
   \int_{-\infty}^{\infty}dx \big(f(x+\omega)-f(x)\big)   H_1(x+\omega,x)\nonumber\\
 &-f(x)\big(H_2(x+\omega,x)+H_2^*(x,x-\omega)\big),
 \label{eq:Sigma_Num}
\end{eqnarray}
where $D_0=ne^2/m$ is the normal-state Drude weight and we used the Fermi-Dirac distribution $f(E)=1/\left(1+e^{E/(k_B T)}\right)$. Notice also that Eq.~(\ref{eq:Sigma_Num}) assumes $\hbar=1$. As shown in the limit $T=0$~ K, the spectroscopically gapless state is present for arbitrary $\Gamma>0$ \cite{Herman21}.

\section{Results}
\label{results}

\subsection{Voigt geometry}
\label{subsec:2}

\begin{figure}[hb]
  \includegraphics[width=0.48\textwidth]{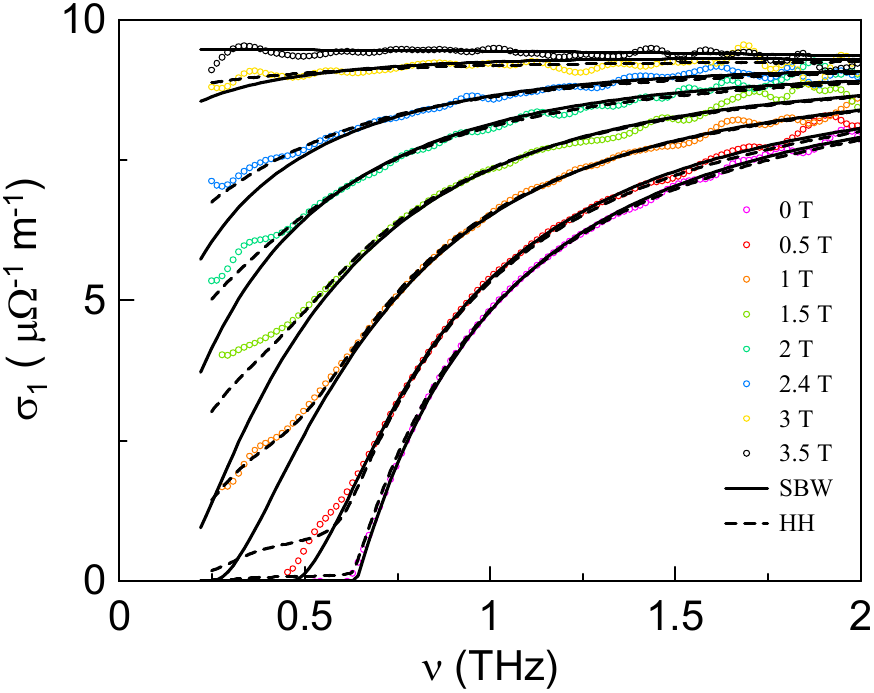}
\caption{Real part of complex conductivity $\sigma_1(\nu)$ of a 58 nm thick Nb film in Voigt geometry measured by Lee \textit{et al}.~\cite{Lee2023}.
Experimental data (points) are theoretically described by the SBW model (solid lines) and the HH model (dashed lines). The imaginary part $\sigma_2(\nu)$ is not shown here. }
\label{fig:2}       
\end{figure}

 In the Voigt geometry, the in-plane external magnetic field fully penetrates thin superconducting films, allowing us to study the effects of a homogeneous magnetic field on their superconducting properties.
Here, we partly revise the work of Lee \textit{et al.}~\cite{Lee2023}. The authors measured the complex conductivity of a $d=58$~nm thick Nb film with a critical temperature of $T_c = 8$~K, see Fig~\ref{fig:2}. From the imaginary part, the London penetration depth $\lambda_L$ was estimated as 0.1~$\mu$m, thus the film was thin compared with its penetration depth, i.e. $\lambda_L \gg d$. The magnetic field averaged over the film thickness $d$ can be approximated by $H_{avg}\approx H_0 (1-d^2/(12 \lambda_L^2))$, which confirms, for the given parameters, a very homogeneous internal magnetic field. In fact, such a Nb film is too thin to develop a sufficiently thick layer of screening currents able to screen the magnetic field.

 \begin{figure}[h]
  \includegraphics[width=0.49\textwidth]{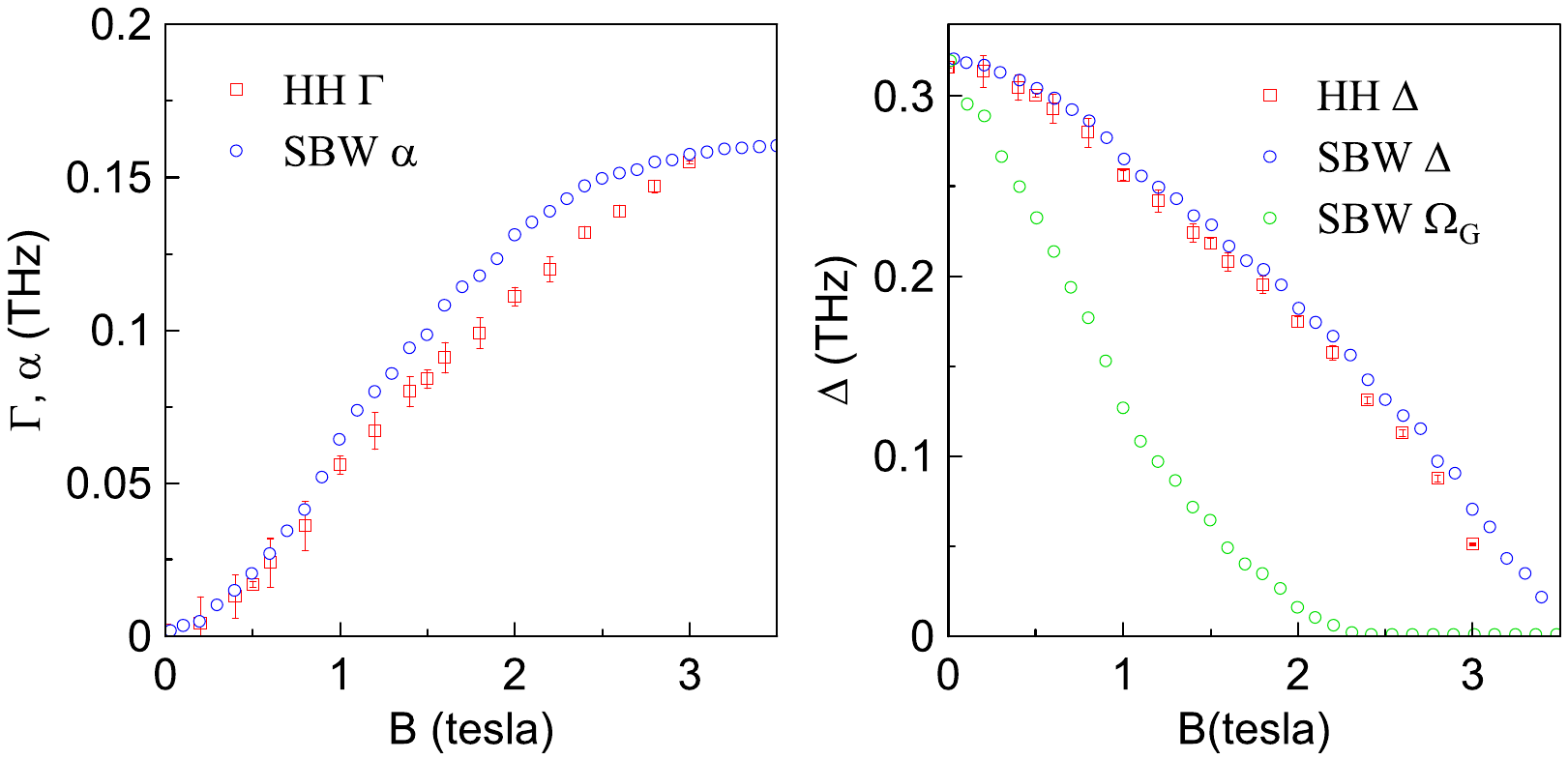}
\caption{Magnetic field dependence of $\Gamma$ and $\alpha$ (left) and $\Delta$ and $\Omega_G$ (right) determined from fits of the experimental data by Lee\textit{ et al}.~\cite{Lee2023}. Circles: SBW model, squares: HH model. }
\label{gap}       
\end{figure}

\begin{figure}[h]
\centering
  \includegraphics[width=0.45\textwidth]{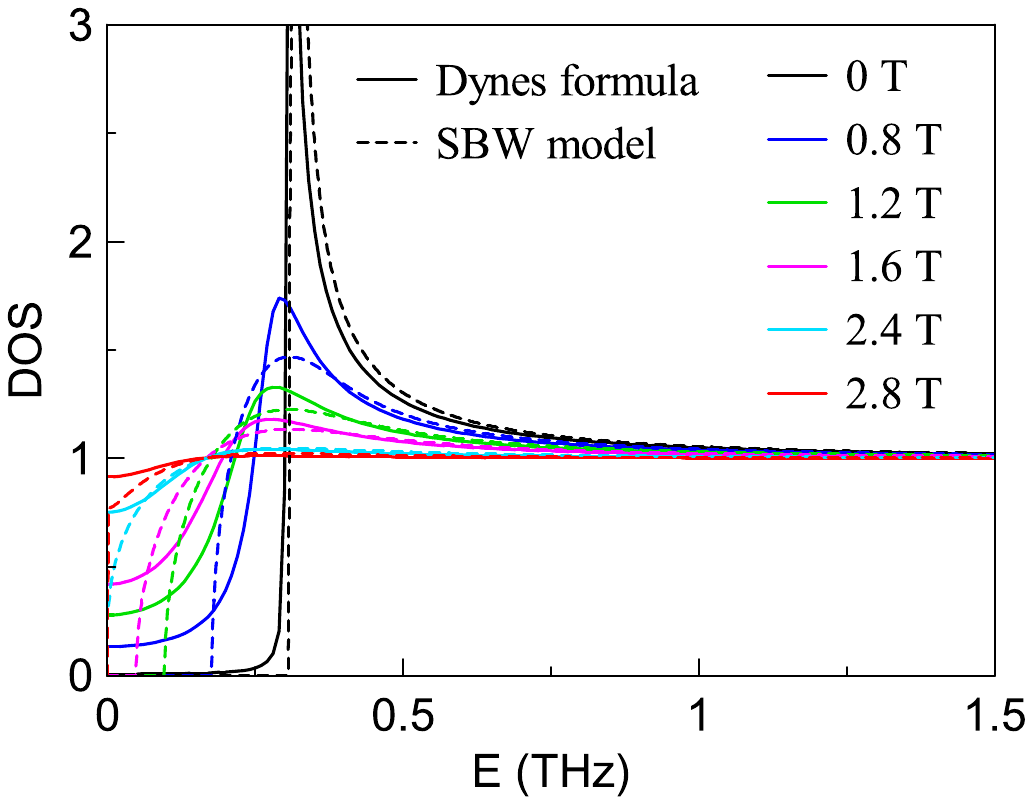}
\caption{Density of states (DOS) in the vicinity of the Fermi energy normalized to the DOS in the normal state for magnetic fields up to  2.8~T. The DOS of the HH and the SBW models are slightly different even for B=0 due to the slight difference in the zero field fits. The DOS was calculated with the Dynes formula, Eq.~(\ref{Dynes_DOS}), using parameters from the fit by the HH model (solid lines), and with the SBW model using parameters from the fit of the THz data (dashed lines). }
\label{DOS}       
\end{figure}

 First, the SBW model was used to fit the zero-field experimental data using the zero-field gap $\Delta_0$, the DC conductivity $\sigma_0$, and the pair-breaking parameter $\alpha$ as free parameters. 
 In magnetic field, $\alpha(B)$ is the only free parameter of the fit since other variables are not field-dependent. The field dependences of the pair-correlation gap $\Delta(B)$ and of the spectroscopic gap $\Omega_{G}(B)$ were determined from the values of $\alpha(B)$ within the SBW model~\cite{Skalski1964}.
Measurements from zero to 3.6~T revealed the suppression of the pair-correlation gap $\Delta(B)$ and of the spectroscopic gap $\Omega_G(B)$. 
Above 2.4~T, $\Omega_G=0$ but $\Delta(B) \neq 0$ and the superconductor enters a gapless state until the superconductivity is completely suppressed at 3.5~T. The SBW model correctly predicts a shift in $\Omega_G$ with the applied magnetic field, but it does not fully describe the shape of the real part of the THz conductivity $\sigma_1(\omega)$ at lower frequencies, see solid lines in Fig~\ref{fig:2}.
We compared the results of the SBW model with the theoretical prediction of Herman and Hlubina~\cite{Herman2017}. First, we fitted the experimental data in a zero magnetic field with DC conductivity $\sigma_0$, the zero-field gap $\Delta_{\Gamma\rightarrow 0}$, the pair-conserving scattering rate $\Gamma_s$, and the pair-breaking scattering rate $\Gamma$ as free parameters. The value of the gap $\Delta$ was approximated by the formula ~\cite{Herman2016, Lebedeva2024a}
\begin{equation}
\Delta=\sqrt{\Delta_{\Gamma\rightarrow0} [\Delta_{\Gamma\rightarrow 0}-2\Gamma]},
\label{gap(gamma)}
\end{equation}
 which is valid in the limit of zero temperature.
For non-zero magnetic fields, we used the resulting values except for the pair-breaking scattering rate $\Gamma(B)$  which was treated as a free parameter. The magnetic-field dependence of the gap $\Delta(B)$ is governed by $\Gamma(B)$ via Eq.~(\ref{gap(gamma)}).
 The real part of the conductivity $\sigma_1(\nu)$ predicted by the HH model, see dashed lines in Fig.~\ref{fig:2},  decreases more slowly at low frequencies due to the presence of the sub-gap absorption \cite{Herman2017}. For low magnetic fields, the SBW model seems to be more accurate, but from 1 T, the HH model describes the experimental results better.
  In Fig.~\ref{gap}, we compare the values of the parameters of the SBW and of the HH theoretical models. The pair-correlation gap $\Delta$ and $\Gamma$, as well as $\alpha$, have qualitatively and quantitatively the same magnetic field dependence. To better understand the difference between these models, we plot the density of states within the SBW model and compare it to the Dynes formula justified by the HH model. The parameters for the DOS are taken from the fits of the complex conductivity $\sigma(\nu)$  with each model.
  They are qualitatively different. Whereas the SBW DOS has no states for $|E|<\Omega_G$ and $B<2.4$ T, the DOS calculated with the Dynes formula using the parameters obtained from the fit by the HH model fills the gap gradually, see Fig.~\ref{DOS}.

\subsection{Faraday geometry}
\label{subsec:1}

 We measured the optical response of three NbN films of different thicknesses (5.3, 11.5, and 30.1 nm) using time-domain THz spectroscopy. In brief, the samples were placed in a He-bath optical cryostat (Spectromag, Oxford Inst.) comprising a superconducting magnet, and their complex THz transmittance was measured using broadband THz pulses. A dedicated procedure was applied to determine the conductivity spectra of the films, whose thicknesses are about five orders of magnitude lower than those of the substrates. Further details of the experimental setup, as well as of the treatment used to extract the THz conductivity spectra from our data, are reported in detail in \cite{Sindler2022}. 

In the Faraday geometry, the magnetic field is perpendicular to the plane of the film, and consequently, so are the vortices. Their presence substantially alters the zero-field properties in type-II superconductors. To interpret the experimental results, we use a model in which the vortices are approximated by normal-state cylinders surrounded by a superconducting matrix. For fields reported here ($\mu_0H>1$\,T), the distance between vortices is smaller than the penetration depth $\lambda_L$, so the magnetic flux penetrates into the superconducting matrix.
The spectral response of the superconducting matrix is described by the HH model~\cite{Herman2017} which incorporates the pair-breaking effects of the magnetic field through the parameter $\Gamma(B)$. Consequently, the gap is modified due to $\Gamma$ according to Eq.~(\ref{gap(gamma)}). 

 \begin{table}[t]
 \caption{Parameters of the measured NbN films---$d_f$: film thickness, $T_c$: critical temperature, $\Gamma_s$: field-independent value of pair-conserving scattering rate, $\Delta_{\Gamma\rightarrow0}$: energy of the gap corresponding to $\Gamma\rightarrow0$, and $\sigma_0$: DC conductivity} 
\label{vzorky}
\begin{tabular}{lccccc}
\hline
\hline
 Sample   & $d_\mathrm{f}$ &  $T_\mathrm{c}$ &  $\Gamma_s$ & $\Delta_{\Gamma\rightarrow0}$ & $\sigma_0$\\
  \  & nm  & K &  THz & THz & $\mu\Omega^{-1}$m$^{-1}$\\
\hline
NbN\textcircled{a}MgO &5.3  & 13.9  & 5.9& 0.62 & 1.55\\
NbN\textcircled{a}Si &11.5  & 12  & 5.4 & 0.51& 0.45\\
NbN\textcircled{a}MgO &30.1&  15.5    & 4.4 & 0.65 &2.52\\
\hline
\hline
\end{tabular}
\end{table}
\begin{figure}[h!]
    \includegraphics[width=0.49\textwidth]{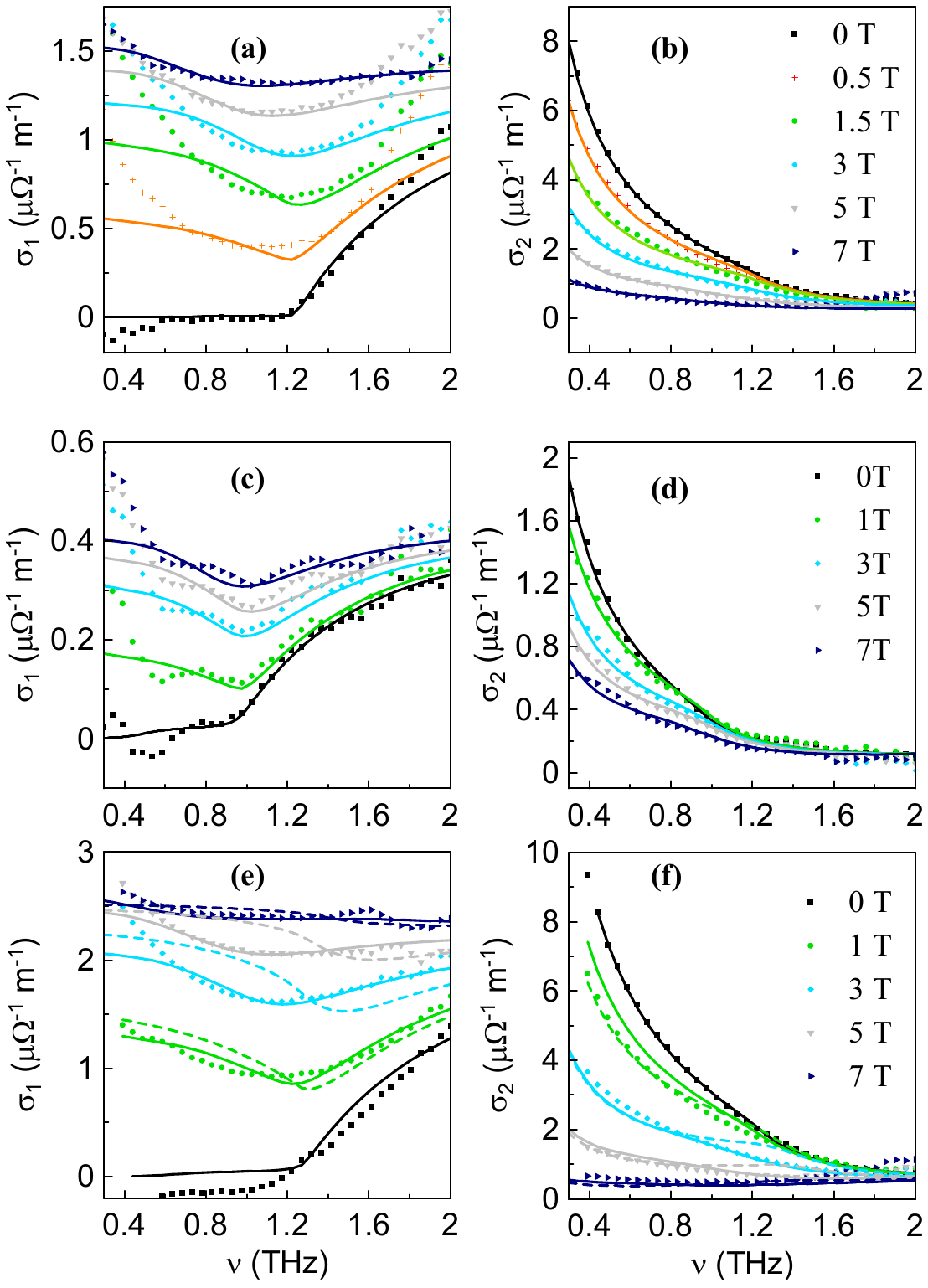}
     \caption{Real (a, c, e) and imaginary (b, d, f) parts of the THz conductivity of NbN thin films measured in the Faraday geometry at 2~K up to 7 T. Three NbN samples of different thicknesses were studied:
         (a-b)  $d_f=5.3$~nm,
         (c-d)  $d_f=11.5$~nm and 
     (e-f) $d_f=30.1$~nm. 
     Symbols: experimental data; full lines: fits using the Maxwell-Garnett theory and the HH model; dashed lines: fits neglecting $\Gamma(B)$.
        }
        \label{s_faraday}
\end{figure}

In the THz range, the conditions for the so-called long-wavelength limit are fulfilled: the wavelengths are much longer than the distances between the vortices, thus the inhomogeneous system can be represented by an effective optical conductivity. 
The geometry of the normal-state inclusions within the superconducting matrix can be well described within the Maxwell-Garnett mo\-del~\cite{Garnett1904} which yields the effective conductivity ${\tilde{\sigma}}_{\rm MG}$:
\begin{equation}\label{mgt}
\frac{\tilde{\sigma}_\mathrm{MG}  - \tilde{\sigma}_{\mathrm{s}}}
{\tilde{\sigma}_\mathrm{MG}  + K \tilde{\sigma}_{\mathrm{s}}}
= f_{\mathrm{n}}  \,
\frac{\tilde{\sigma}_{\mathrm{n}}-\tilde{\sigma}_{\rm{s}}}
{\tilde{\sigma}_{\mathrm{n}}+K\tilde{\sigma}_{\mathrm{s}}},
\end{equation}
  where $f_{\rm n}$ and  ${\tilde{\sigma}}_{\rm n}$ are the volume fraction and the conductivity of the normal-state vortex cores, respectively, and
  ${\tilde{\sigma}}_{\rm s}$ is the conductivity of the superconducting matrix. $K$ is the shape factor of inclusions which was set to 
  $K=1$, corresponding to the value for cylinders with their axes perpendicular to the electric field, as in the present case.
   
\begin{figure}
    \includegraphics[width=0.485\textwidth]{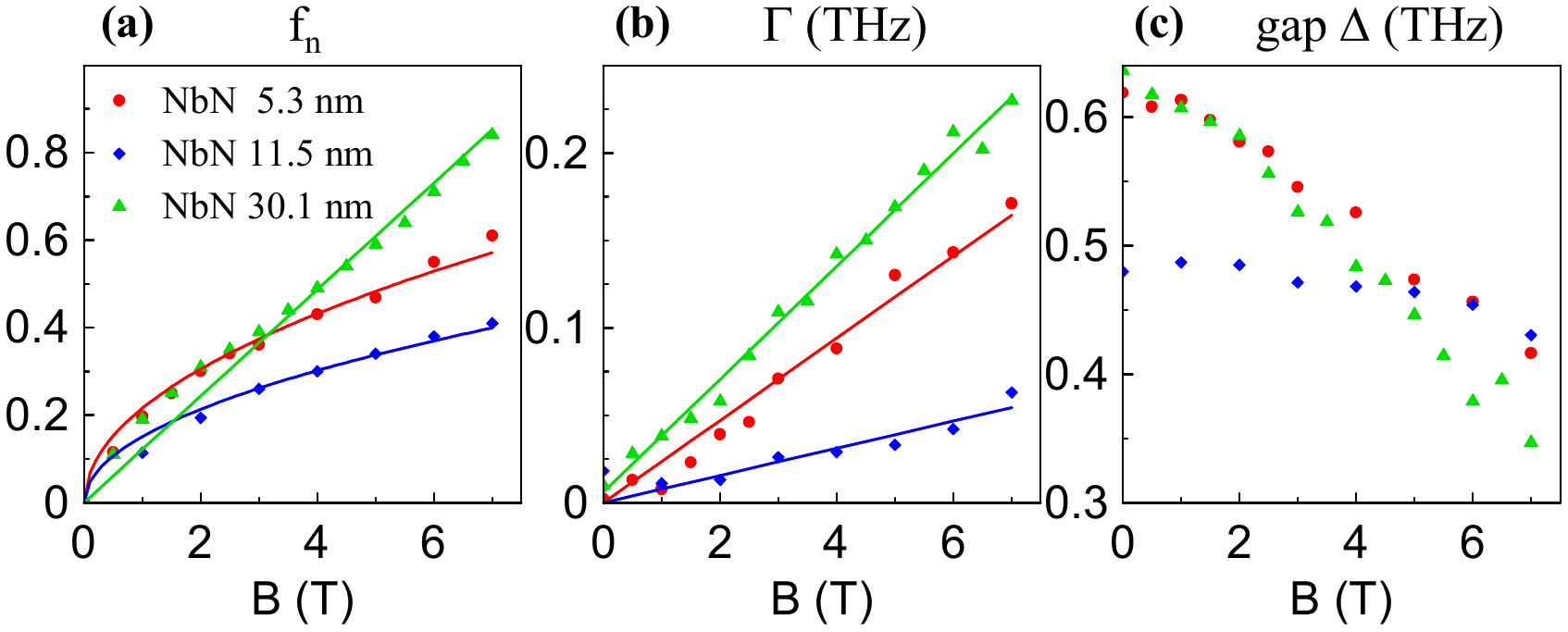}
     \caption{  The normal state volume fraction $f_n$, the pair-breaking rate $\Gamma$ and the superconducting gap $\Delta$ as functions of the magnetic field. Points: values obtained by the fits of the spectra with the HH model, lines: fitted curves. }
    \label{fit_param}
\end{figure}

  The THz conductivity spectra of the three NbN films are shown in Fig.~\ref{s_faraday}. 
     In our fits, we used two free parameters: the volume fraction of the vortex cores $f_{\rm n}$, and the pair-breaking scattering rate $\Gamma$ which influences the shape of the superconducting conductivity spectra $\sigma_s(\nu)$. The remaining parameters are field independent, and their values were taken from the zero-field fit.
   Fits in the Figure 6(e, f) compare the following calculations within the Maxwell-Garnett theory (MGT). Both use HH model for $\sigma_{\mathrm{s}}(\nu)$. However, dashed curves use zero magnetic field $\Gamma(0)$, while solid curves account for magnetic-field dependence through $\Gamma(B)$. The latter demonstrates improved agreement with experimental data.   
   The pair-breaking scattering rate $\Gamma(B)$ increases with the magnetic field for all three NbN films, as shown in Fig.~\ref{fit_param}(b). The gap energy $\Delta$ [shown in Fig.~\ref{fit_param}(c)] was evaluated using Eq.~(\ref{gap(gamma)}) from the fitted values of $\Gamma$.

  The normal fraction should be a linear function of the magnetic field ($f_n=B/B_{c2}$), but this holds only for the thickest sample, see Fig.~\ref{fit_param}(a). The square root dependence $f_n \propto a B^{1/2}$ observed by Xi \textit{et al.}~\cite{Xi2013} successfully describes $f_n(B)$ for the 5 nm and the 11.5 nm thick NbN films. A similar square-root dependence was also observed in overdoped La$_{2-x}$Ce$_x$CuO$_4$~\cite{Tagay2021}.
 The sublinear behaviour can be attributed to the limitations of the simple effective medium model, or possibly 
to the shrinking of the vortex cores~\cite{Sonier2004,Ning2010} with increasing magnetic field.

It is important to note that the presented theoretical description is incomplete without accounting for vortex motion in the Faraday geometry. Presently, no unified model exists that would simultaneously incorporate (i) the BCS theory description of optical conductivity, (ii) the presence of vortex cores, and (iii) the vortex-motion-induced absorption. In particular, the THz regime exhibits a distinct behavior compared to the GHz range: on picosecond timescales, the vortex motion is limited to localized oscillations around pinning sites, and the effective potential barriers may differ from those governing the low-frequency dynamics.

The contribution of vortices to conductivity can be described by the formula presented in the review paper by Pompeo \textit{et al.}~\cite{Pompeo2008} which 
takes the same form for the Coffey-Clem~\cite{CCmodel1991} and Brandt~\cite{Brandt1991} models. However, the models differ in their physical interpretations.
A key parameter is the depinning frequency $\nu_p$, above which the driving force dominates the pinning, leading to dissipative flux flow. An estimate from measurements by Basistha \textit{et al.} \cite{Basistha2024} suggests that the depinning frequency of NbN is in the MHz range, while for niobium, Pompeo et al.~\cite{Pompeo2010} report 
$v_p= 7.2$\,GHz. This confirms that THz waves act deep within the flux-flow regime, where vortices are highly mobile.

In our prior work~\cite{Sindler2014}, we analysed vortex dynamics in the NbN film deposited on Si substrate (the one examined also here) and demonstrated the role of the vortex motion in the upturn of $\sigma_1$ at low THz frequencies which is not accounted by the present model.
 This underscores the necessity of developing an integral description of the vortex dynamics to fully capture the electrodynamic response of superconductors.

\section{Conclusions}
\label{sec:conclusions}

The theoretical model by Herman and Hlubina~\cite{Herman2017} for conventional superconductors allowed us to calculate the optical conductivity for varying ratios of the pair-conserving and pair-breaking processes. We demonstrated that the HH model can well describe the optical conductivity in magnetic fields both in Voigt and Faraday geometries owing to the varying pair-breaking scattering rate~$\Gamma$. In this way, the model serves as an effective phenomenology describing the pair-breaking effect of the increasing external magnetic field which suppresses the $T_c$, fills the gap in the density of states, and affects the optical conductivity. In the Voigt geometry, the HH model offers a better theoretical description of the THz conductivity of the Nb thin film measured by Lee \textit{et al.}~\cite{Lee2023} than the SBW model used previously. In contrast with the SBW model, it predicts the existence of a non-zero DOS at the Fermi energy even for $B=0$ T, as well as the filling of the gap region with an increasing magnetic field. Subgap absorption described by the gapless regime in the HH model better describes measurements of $\sigma_1(\nu)$ at low frequencies. Comparison of the SBW model and HH model based analysis therefore opens an interesting question about the presence of the gapped-superconducting state once $B\neq 0$.

In the Faraday geometry, the THz response of the vortex lattice is described by the Maxwell-Garnett effective medium theory~\cite{Garnett1904} assuming a superconducting matrix following the HH model~\cite{Herman2017}.
As one would expect, the values of resulting parameters fulfill relations $\Gamma \lesssim \Delta \lesssim \Gamma_s/10$, suggesting the possibility of applying the HH model in the dirty limit \cite{Herman2017}. However, the presented formulation allows for an explicit presence of the pair-conserving scattering; this may be useful for analyzing cleaner superconducting systems. 

Last but not least, let us emphasize that the presented analysis utilizes the HH model as an effective phenomenology with the fewest free parameters, taking into account $\Gamma(B)$ as the effective pair-breaking scattering rate. This approach allows for a straightforward comparison with related results reported recently in the literature. The difference in shapes of $\Gamma(B)$ dependences considering different geometries probably originates from a combination of the Zeeman splitting and the spin-orbit interaction in the Voigt geometry and from the orbital effects in the Faraday geometry. To proceed further with the analysis of the presented experimental data, and to better distinguish the consequences of individual effects on the THz conductivity measurements, one would need a more elaborate description of the Zeeman splitting and of the spin-orbit interaction, to take into account the screening current generation (orbital effects), as well as the role of intrinsic pair-breaking and pair-conserving disorder. In the best-case scenario, such an improved model could go beyond the Born approximation limit, allowing for higher magnitudes of the individual effects.

\section*{Acknowledgements}
We are grateful to K. Ilin and M. Siegel for preparing and characterizing the NbN sample,
as well as to J. H. Kim and J. E. Lee who were very kind to provide their data~\cite{Lee2023}. We are also thankful to R. Hlubina and L. Skrbek for discussions.
We acknowledge
the financial support from the Czech Science Foundation
(Project No. 21-11089S), by the European Union and the Czech Ministry of Education, Youth and Sports (Project TERAFIT - CZ.02.01.01/00/22\_008/0004594) and by the INTER-COST (Project LUC24098).
This work was also supported by the Slovak Research and Development Agency under the Contract no. APVV-23-0515 and by the European Union’s Horizon 2020 research and innovation programme under the Marie Skłodowska-Curie Grant Agreement No.~945478.

\bibliography{Dynes}

\end{document}